# Efficient photogeneration of charge carriers in silicon nanowires with a radial doping gradient

D.H.K. Murthy[1,7], T. Xu[2], W. H. Chen,[3] A.J. Houtepen[1], T.J. Savenije[1], L.D.A. Siebbeles[1], J. P. Nys[2], C. Krzeminski[2], B. Grandidier[2*], D. Stiévenard[2], P. Pareige[3], F. Jomard[4,] G. Patriarche[5], O. I. Lebedev[6]

[1]Optoelectronic Materials Section, Department of Chemical Engineering, Delft University of Technology, 2628 BL Delft, The Netherlands.

[2] Institut d'Electronique, de Microélectronique et de Nanotechnologie, IEMN, (CNRS, UMR 8520, Département ISEN, 41 bd Vauban, 59046 Lille Cedex, France.

[3] Groupe de Physique des Matériaux, Université et INSA de Rouen, UMR CNRS 6634, Av. de l'université, BP 12, 76801 Saint Etienne du Rouvray, France.

[4] GEMAC, Université de Versailles-Saint-Quentin, UMR CNRS 8535, 1 place Aristide Briand, 92125 Meudon, France.

[5] CNRS-Laboratoire de Photonique et de Nanostructures, Route de Nozay, 91460 Marcoussis, France.

[6] Laboratoire CRISMAT. UMR 6508, CNRS-ENSICAEN, Université de Caen, 6Bd Marechal Juin 14050 Caen, France.

[7] Dutch Polymer Institute (DPI), P.O. Box 902,5600 AX Eindhoven, The Netherlands.

**ABSTRACT** From electrodeless time-resolved microwave conductivity measurements, the efficiency of charge carrier generation, their mobility, and decay kinetics on photo-excitation were studied in arrays of Si nanowires grown by the vapor-liquid-solid mechanism. A large enhancement in the magnitude of the photoconductance and charge carrier lifetime are found depending on the incorporation of impurities during the growth. They are explained by the internal electric field that builds up, due to a higher doped sidewalls, as revealed by detailed analysis of the nanowire morphology and chemical composition.

*bruno.grandidier@isen.iemn.univ-lille1.fr

Thin films consisting of arrays of semiconductor nanostructures exhibit electronic, optical and thermal properties that derive to a large extent from the periodic arrangement of their nanometer scale constituents. For example, arrays of Si wires have been shown to successfully increase light-trapping compared to planar semiconductor thin films [i,ii], while providing a better geometry to collect photoexcited carriers [iii,iv]. Periodic nanoholes etched in thin films of silicon modify the mean free paths of phonons and greatly reduce the thermal conductivity of the film [v]. As a result, arrays of semiconductor nanostructures offer the prospect to improve the energy conversion efficiency in devices such as solar cells or Peltier elements.

Whatever the foreseen application is, the conversion efficiency partly depends on the transport properties of the array. Although these properties have been investigated for single semiconductor nanowires (NWs) [vi] their electrical conductivity may vary from one nanowire to another in the same array depending on slight variations in the NW morphology such as its diameter or its surface roughness. Key transport parameters, like the carrier lifetime and mobility, are still not well known and are usually assumed to be similar to their bulk counterparts in order to determine the conversion efficiency of the NWs [vii]. In addition, difficulties in establishing Ohmic electrical contacts often complicate interpretation of the measurements [viii,ix,x]. Therefore, new experimental tools are required to investigate the transport properties of semiconductor nanostructure ensembles.

As an answer to this challenge, we show that the time-resolved microwave conductivity (TRMC) technique is a promising method to measure, without any electrode, the transient photoconductance of Si NW arrays of semiconductor nanostructures [xi]. As a prototypical example, we investigated the photoconductive properties of arrays of Si NWs grown by the vapor-liquid-solid (VLS) technique using gold as catalyst. Depending on the incorporation of impurities, the TRMC technique allows to measure strong differences in the efficiency of the charge carrier photogeneration, their mobility, and decay over time on photoexcitation. Such differences are rationalized from the detailed knowledge of the NW structural and chemical composition, obtained using transmission electron microscopy (TEM) and atom probe tomography (APT). They show how formation of an internal electric field due to differences in radial doping can be advantageously used to separate photoexcited electron hole pairs.

Si NWs were prepared by the VLS method from silane on quartz substrates using the chemical vapor deposition (CVD) technique. Gold nanoparticles were used as catalysts. They were formed at 700°C in the CVD chamber from the deposition of a gold film with a thickness of 2 nm. The growth was performed with a silane partial pressure of 1 mbar, a total pressure in the chamber of 10 mbar and a temperature of 430°C. N-type doped Si NWs were also grown with a silane flow rate of 300 sccm (standard cubic centimeters per minute), whereas the flow rate of phosphine diluted in $H_2$ was 1 or 5 sccm. The morphology of the Si NWs was characterized by scanning (SEM) and transmission (TEM) electron microscopies. Their chemical composition was obtained with the atom probe tomography (APT) measurements, that was conducted with a laser-assisted atom probe tomography (CAMECA LAWATAP). In contrast to the previous method that we developed for specimen preparation [xii], Si NWs were directly manipulated on their substrate and welded from the top of the NW onto a pre-prepared W supported tip in a Cross Beam (Focused Ion Beam - Gas Injection System- SEM) workstation. Typically, when NWs are taken from the substrate, they break up in a part of the shaft that is far from both ends. The analysis of the chemical composition that takes place in this region over several hundreds of nm along the growth axis is thus representative of the chemical composition of the rest of the NWs, with the exception of the extremities, where the growth conditions are different. The orientation of the NWs on the tip can be controlled by changing the W tip vertex angle and the welding position. As The specimen analyzed by APT should have good mechanical properties, $(Me_3)MeCpPt$ and Pt are chosen as the deposition precursor and the solder for reinforcement. The APT characterization was performed using a UV laser as the pulse source (wavelength of 343 nm, power of 4 mW). The chamber pressure and specimen temperature were maintained at $2\times10^{-10}$ mbar and 80 K respectively. The detection limit of the phosphorus atom is $1.2\times10^{18}$ at.cm$^{-3}$.

A detailed description of the TRMC technique can be found elsewhere [xiii,xiv] . In brief, Si NWs grown on quartz substrates were placed in an X-band (8.4 GHz) microwave cell and photoexcited with a linearly polarized laser pulse from an Optical Parametric Oscillator pumped at 355 nm with the third harmonic of a Q-switched Nd:YAG laser (Vibrant, Opotek). Photogeneration of mobile charge carriers in the sample leads to an increase of the conductance, $\Delta G(t)$, and consequently to the enhanced absorption of microwave power by the sample. The time-dependent change of the conductance is obtained from the normalized change in reflected microwave power $(\Delta P(t)/P)$ from the cell, according to [Ref. **Erreur ! Signet non défini.**,**Erreur ! Signet non défini.**,xv,xvi]:

$$\frac{\Delta P(t)}{P} = -K\Delta G(t) \tag{1}$$

The dimensions of the microwave cell and the geometrical and properties of the media in the microwave cell determine the sensitivity factor, K (see supporting information). For equation 1 to be applicable $\Delta P/P$ is kept below $5 \times 10^{-3}$ since higher values might change the microwave field pattern in the cell. The absolute value of the conductance is related to the number of negative (electrons) and positive (holes) charge carriers and their mobilities. Combining the maximum change in conductance ($\Delta G_{max}$) with the incident light leads to the following equation:

$$\eta\Sigma\mu = \frac{\Delta G_{max}}{\beta e J_0} \tag{2}$$

in which the parameter $\eta\Sigma\mu$ denotes the product of the charge carrier generation yield ($\eta$) per incident photon and the sum of the electron and hole mobilities ($\Sigma\mu$). $\beta$ is the ratio between the broad and narrow inner dimensions of the waveguide used, $e$ is the elementary charge, $J_0$ is the laser fluence i.e. the number of incident photons per unit area per pulse.

Figure 1 shows typical images of the Si NWs grown on quartz substrates obtained with SEM and TEM. A high density of Si NWs is observed. Their sidewalls are straight and parallel to the growth axis, as it is seen in the high resolution TEM (HRTEM) image of Fig. 1(b). When the growth takes place with the addition of phosphine, the overall morphology of the NWs does not change. For both types of NWs, the histograms of the NW diameter distribution reveals, in Fig. 1(d), that doping the NWs do not modify their diameter, the mean diameter being 65 nm for doped NWs and 66 nm for non doped NWs (the diameter is measured just below the Au seed particle). Finally, we note that the NWs are slightly tapered. From the high-angle annular dark field (HAADF) scanning transmission electron microscopy (STEM) image of the n-type Si NW shown in Fig. 1(c), we measured a variation of 21 nm between the base and the top of the NW, indicating that the lateral growth rate is about 140 times slower than the axial growth rate. Despite this lateral deposition, the sidewalls do not appear defective. Indeed, HRTEM images of the Si NWs, such as the one presented in Fig. 1(c), show that the cristallographic structure of the core is still preserved at the surface of the NW.

First, TRMC experiments were performed on non-doped and n-doped Si NWS. While the amplitude of the photoconductance signal was weak on photoexciting the samples through the quartz substrate, more than 10 times higher signals could be measured when the the NWs were directly photoexcited, as shown in Figure 2(a). This difference is explained as follows. The growth of Si NWs leads to the formation of a very thin layer of amorphous Si between the Si NWs, on the top of the quartz substrate. This layer contains a lot of Au impurities that can act as traps for the charge carriers, leading to the formation of a disordered layer with numerous point defects in contrast to the top part of the sample, where the Si NWs are crystalline, as demonstrated by the TEM images. Such variation in the structural quality between the top and the bottom parts of the structure grown on the quartz substrate accounts for the significant change in the measured amplitude of the photoconductance.

Focussing on the photoconductance traces obtained for the top part of the samples only, a constant mean diameter of the Si NWs between the doped and non-doped wires (see Fig. 1(d)) ensures similar optical absorptions for both types of arrays [xvii]. As a result, the magnitude and lifetimes measured on the photoconductance traces can be compared between the different samples. On formation of mobile charges due to the nanosecond laser pulse, the TRMC signal increases during the first few nanoseconds, as shown in Fig. 2(b). Note that, since no electrodes are applied, the decay of the TRMC signal is solely due to recombination and/or trapping of charges. Clearly, n-doping leads to a substantial enhancement of $\Delta G_{max}$. In addition, a longer half-lifetime of the charges ($\tau_{1/2} = 42 \pm 2$ ns) as compared to the charges generated in the non-doped wires ($\tau_{1/2} = 6.5 \pm 2$ ns) is found.

Si NWs are prone to have many intraband gap surface states and dangling bonds [xviii,xix]. N-type doping of Si NWs is therefore expected to lead to the population of these surface states with electrons [xx]. Hence the outer surface of n-type doped Si NWs carries excess negative charge that is expected to influence the band bending at the sidewalls of the NW [xxi,xxii]. In order to unravel if the beneficial effects of doping on the photoconductance are not due to this charge trapping effect, an additional etching treatment was applied. Etching with HF removes the oxide layer on the surface that ultimately minimizes the defect density and/or dangling bonds by a factor of 50 as reported earlier [**Erreur ! Signet non défini.**]. As a consequence, the charge trapping effect at the surface of the NW is largely reduced. Note that the TRMC experiments were performed immediately after etching in the absence of oxygen to avoid re-oxidation of these wires. As shown in Figure 2(c) the magnitude of photoconductance is increased by a factor of less than two after etching as compared to the value before etching. For the 5 sccm doped NWs a similar observation was made. Most strikingly, the photoconductance decays are hardly affected by the etching process, indicating that charge trapping by intra band-gap surface states on the Si NW sidewall has a negligible effect on the lifetime of the photoexcited carriers.

In order to explain the measured recombination lifetimes between non-doped and n-type doped Si NWs, an analysis of the NW chemical composition was performed. The chemical composition of the n-type doped Si NWs was investigated with the APT and secondary ion mass spectrometry (SIMS) techniques. Figure 3(a) shows a selected volume reconstruction of the core of a doped Si NW, where phosphorus impurities are clearly detected. The concentration of phosphorus impurities establishes at $2\times10^{18}$ at.cm$^{-3}$ in the center of the Si NWs and gradually increases towards the surface, in agreement with previous reports [xxiii,xxiv]. At the surface of the Si NWs, the doping level may be different since the incorporation rate of impurities on the NW sidewall has been previously found to be different when lateral growth occurs [xxv]. A comparison with the SIMS experiments performed on a film grown simultaneously with the NWs reveals that the film thickness is similar to the overgrown layer on the sidewall and that the phosphorus impurity concentration in the film amounts to $5\times10^{19}$ at.cm$^{-3}$. Therefore we expect the shell of the n-type doped Si NWs to have a similar concentration of phosphorus impurities, which is an order of magnitude higher than the one found in the core of the NWs. Such an inhomogeneous radial distribution of the dopants suggests a variation of the position of the Fermi level with respect to the edge of the bands in the radial direction.

SEM and TEM images also show bright protrusions at the top part of the Si NWs. The protrusions are attributed to Au-rich clusters, because Au is known to diffuse from the seed particle at the end of the growth due to the reduction of the silane partial pressure, while the temperature is still high in the CVD chamber [xxvi,xxvii,xxviii]. As this region only extends over 200-300 nm below the seed particles as pointed out by the brackets in Fig. 1(a), it is quite small in comparison with the rest of the shaft (see Fig. 1(c)) and its contribution in the photoconductance can thus be neglected. From the knowledge of the chemical composition of the NWs, the following model arises: the radial variation of the phosphorus impurities in the Si NWs leads to the formation of a n-n+ core-shell junction. Therefore, the bands are not flat across the NW diameter and an internal electric field is built up.

The profile of the radial phosphorus distribution in the doped Si NWs was used to obtain the internal electric field in the NWs by solving the 2D Poisson equation (DESSIS - Release Z.2007.03, Synopsys Co., Ltd). We consider that they are covered with a thin oxide layer, with the Fermi level pinned midgap at the surface of the NWs. We also presume that all the impurities are electrically active. Figure 4 shows the variation of the electric field across the diameter of a n-type doped Si NW. For comparison, the dopant concentration is also drawn in the inset of Fig. 4. An abrupt variation of the electrical field is found at the interface between the $n^+$ doped shell and the n doped core. Then, there is almost no electrical field in the rest part of the shell, due to the high doping level in this region, whereas the electric field varies smoothly in the core to reach zero at the center of the NW. The existence of an internal electric field allows separating the photoexcited carriers. The electrons get preferentially trapped in the highly doped shell, whereas the holes drift towards the core. The recombination lifetime thus becomes longer in comparison with the non doped Si NWs, because recombination is hindered by the spatial separation of the carriers.

The presence of the electric field also accounts for the negligible effect of the chemical composition of the NW sidewall on the charge carrier lifetime as demonstrated in the inset of Figure 2(c). Indeed, holes are repelled from the surface avoiding recombination with electrons through surface states. Such a result is confirmed by studying the product of the quantum yield ($\eta$) and the total mobility ($\Sigma\mu$) as a function of the laser fluence. Figure 5 depicts the variation of this product before and after etching the sidewalls of the Si NWs. While the maximum photoconductance increases with laser pump fluence, $J_0$, according to $\Delta G_{max} \propto J_0^{\alpha}$, the product $\eta\Sigma\mu$ is proportional to $J_0^{\alpha-1}$. For the doped Si NWs the value of $\alpha$ amounts to 0.82 and 0.80 before and after etching, respectively. The fact that the slopes do not change substantially on etching ($\alpha$ values are very similar) implies that recombination on fast time scales (< 1 ns) is not mediated by the intraband-gap surface states. This observation renders support to our previous conclusion that recombination takes place in the core of the NWs.

Interestingly, for all fluences studied the n-doped Si NWs show an order of magnitude higher $\eta\Sigma\mu$ values as compared to non-doped Si NWs. If only first order decay processes occur, $\eta\Sigma\mu$ is constant and $\alpha = 1$ [**Erreur ! Signet non défini.**,xxix]. If also higher order decay processes take place such as bimoleculer recombination, $\alpha < 1$. For the non-doped Si NWs the slope $\alpha$ amounts to 0.90 and 0.91 before and after etching. The difference in slopes between the non doped ($\alpha = 0.90$) and n-type doped Si NWs ($\alpha = 0.82$) indicates that recombination processes in both types of NWs are different. On assuming $\eta$ to be 1, the minimum mobility value for 1 sccm NWs at the lowest incident intensity is then found to be 8 and 16 $cm^2$/Vs before and after etching, respectively. For NWs grown with a phosphine flow rate of 5 sccm, similar values are found. Although no discrimination can be made between the individual contributions from positive or negative charge carriers to the total mobility in using the TRMC technique, we can compare the measured AC mobility to the DC hole and electron mobilities in bulk crystalline silicon for similar dopant concentrations. These mobilities are expected to be around 100 $cm^2$/Vs for electrons in the shell and 5 $cm^2$/Vs for holes in the core [xxx,xxxi,xxxii]. The measured mobility is thus consistent with the DC hole mobility in n-type Si bulk. Although the total mobility could be higher if the actual yield of charge carriers on the time scale of the TRMC experiment is less than 1, we rather suspect that the electron mobility in the Si NWs is small or even negligible. Indeed, due to the internal electric field, electrons drift to the shell. As the NWs stand predominantly in an upright position, electrons oscillate in the shell along the radial direction due to the high frequency electric field component of the microwaves. But, because of a limited shell thickness, that does not exceed 10 nm at the base of the NWs and is smaller towards the top, electrons are likely to be constantly scattered at the boundaries of the shell. As a result their mobility is significantly lowered, suggesting that only holes contribute to the measured mobility.

In summary, we have investigated the photoconductance of Si NWs with the contactless timre-resolved microwave conductance technique. The lateral overgrowth of a highly doped thin layer leads to the formation of core-shell NWs, where the internal electric field efficiently separates photoexcited electron-hole pairs. While a deep understanding of the photoconductive properties of the Si NWs requires their analysis with sophisticated complimentary characterization techniques such as TEM and APT, key parameters can be routinely measured using the non-invasive TRMC technique. Due to its simplicity, this technique can be applied to a wide range of nanostructure ensembles and offers the prospect to perform in-situ analyses.

**Acknowledgment.** The work of DHKM forms the research program of the Dutch Polymer Institute (DPI, project #681). We acknowledge Daniela Ullien and Louis de Smet from the section Nano-Organic Chemistry of Chemical Engineering department at TU Delft for assistance in etching with HF. A.J.H. acknowledges the 3TU Centre for Sustainable

Energy Technologies (Federation of the Three Universities of Technology) for financial support. The authors also acknowledge financial support form the Délégation Générale pour l'Armement under contract REI-N02008.34.0031.

FIGURE CAPTIONS

**FIGURE 1.** (a) SEM image of high-density Si NWs grown by CVD on a quartz substrate with a high silane partial pressure of 1 mbar at 430°C. The brackets point to the upper area of two NWs that are covered with Au-rich clusters due to the decrease of the silane pressure at the end of the growth. (b) HRTEM image of such a Si NW that shows the straigth sidewalls and the absence of Au clusters. (c) HAADF-STEM image of n-type doped Si NW grown at high silane partial pressure, where the tapered shape is better seen by comparing the diameter of its top (right inset) and bottom (middle inset) ends. (d) Histograms for the distribution in non-doped and n-type doped Si NW diameter.

**FIGURE 2.** (a) Photoconductance transients observed for doped Si NWs when the excitation is performed from the front (FS) or the back (BS) side of the sample. The photoexcitation was set at 420 nm using an incident intensity of 0.65 $\mu J/cm^2$.pulse. (b) Photoconductance transients observed for non-doped and n-doped Si NWs photoexcited at 420 nm using an incident intensity of $55 \times 10^{-9}$ $J/cm^2$.pulse for different flow rates of phosphine diluted in $H_2$. The flow rate is expressed in units of standard cubic centimeters per minute (sccm). (c) Dependence of the photoconductance transients on the surface chemistry for n-doped Si NWs grown with 1 sccm of phosphine diluted in $H_2$. The excitation was set at 420 nm using an incident intensity of $55 \times 10^{-9}$ $J/cm^2$.pulse. All the insets show corresponding normalized photoconductance transients.

**FIGURE 3.** (a) End-on view and side view of the atomically resolved three-dimensional phosphorus distribution in the core of a 70 nm–diameter Si NW. Phosphorus atoms appear as pink dots, whereas the black matrix represents Si atoms. (b) Average phosphorus concentration along the radial axis starting from the middle of the NW towards the edge. Inset: SIMS analysis of the phosphorus concentration in the 25 nm-thick amorphous Si film grown on a $SiO_2$ surface during the growth of the P-doped Si NWs.

**FIGURE 4**. Calculated radial distribution of the electrical field in a n-type doped Si NWs grown, based on the  phosphorus doping profile shown in the inset.

**FIGURE 5.** $\eta\Sigma\mu$ values as a function of incident light intensity ($J_0$) for doped and non-doped Si NWs at 420 nm. Left and right panel corresponds to (a) before and (b) after etching with HF, respectively. Solid lines are fits to data points using equation $\eta\Sigma\mu \propto J_0^{\alpha-1}$

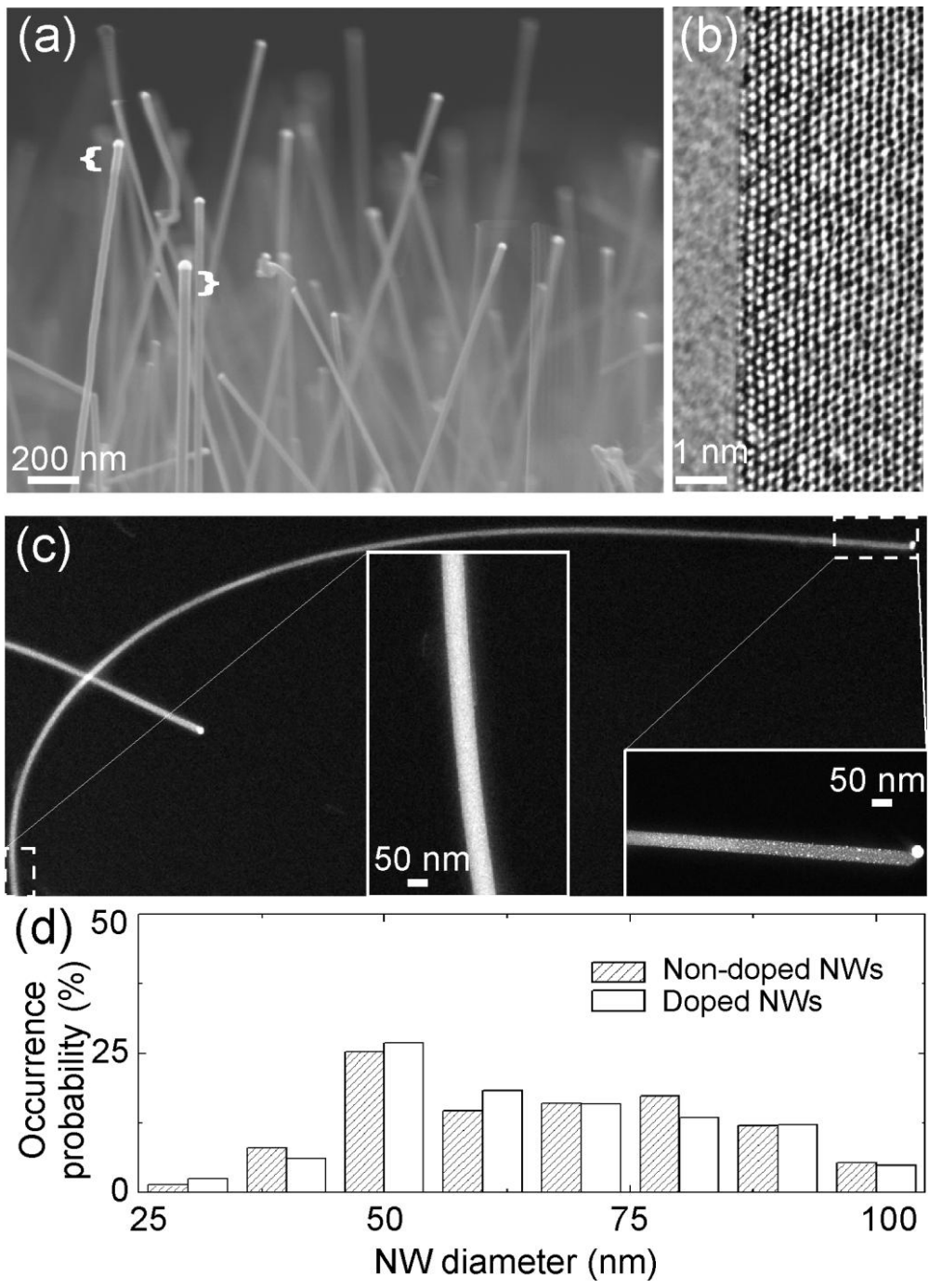

Figure 1

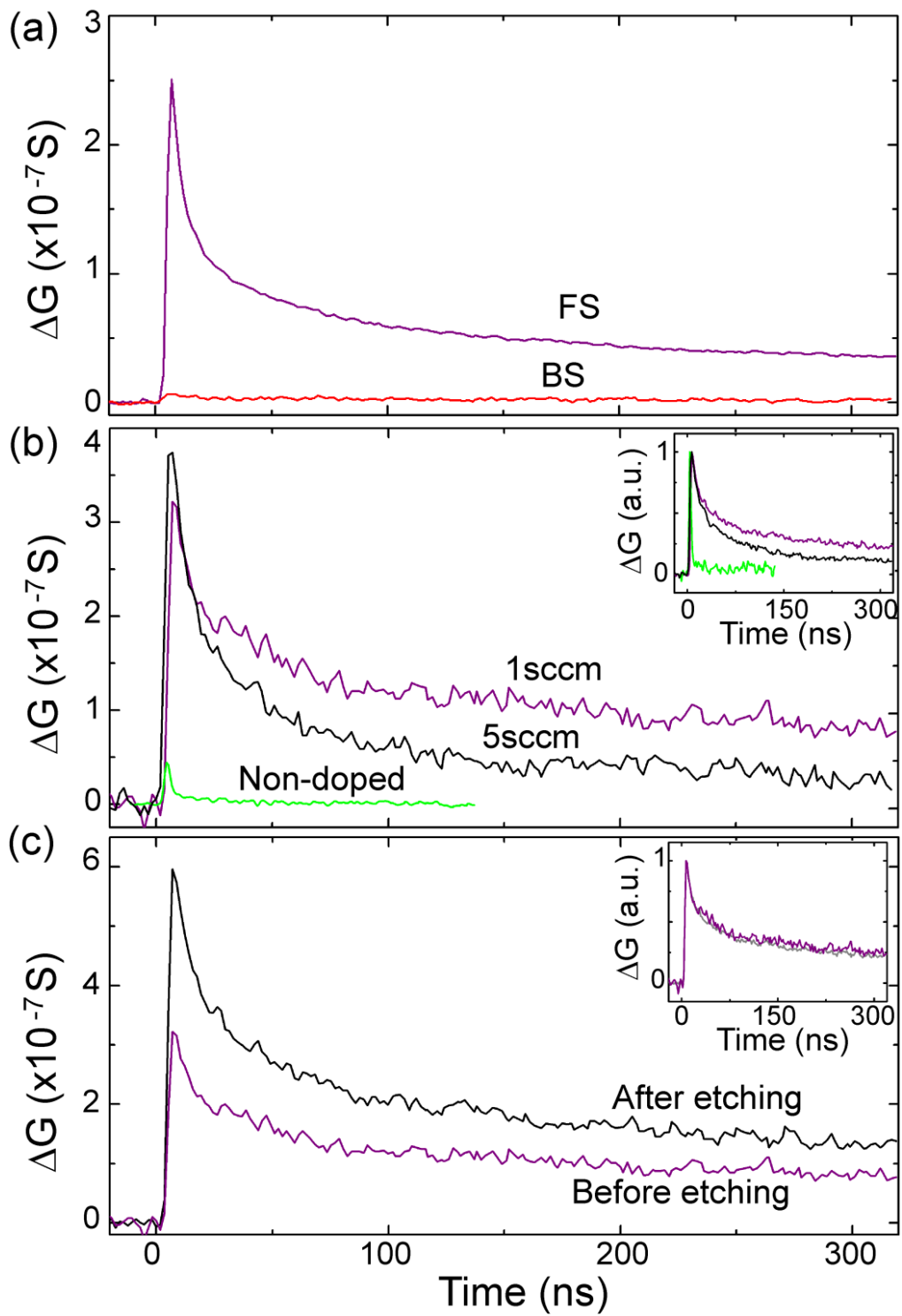

Figure 2

(a)

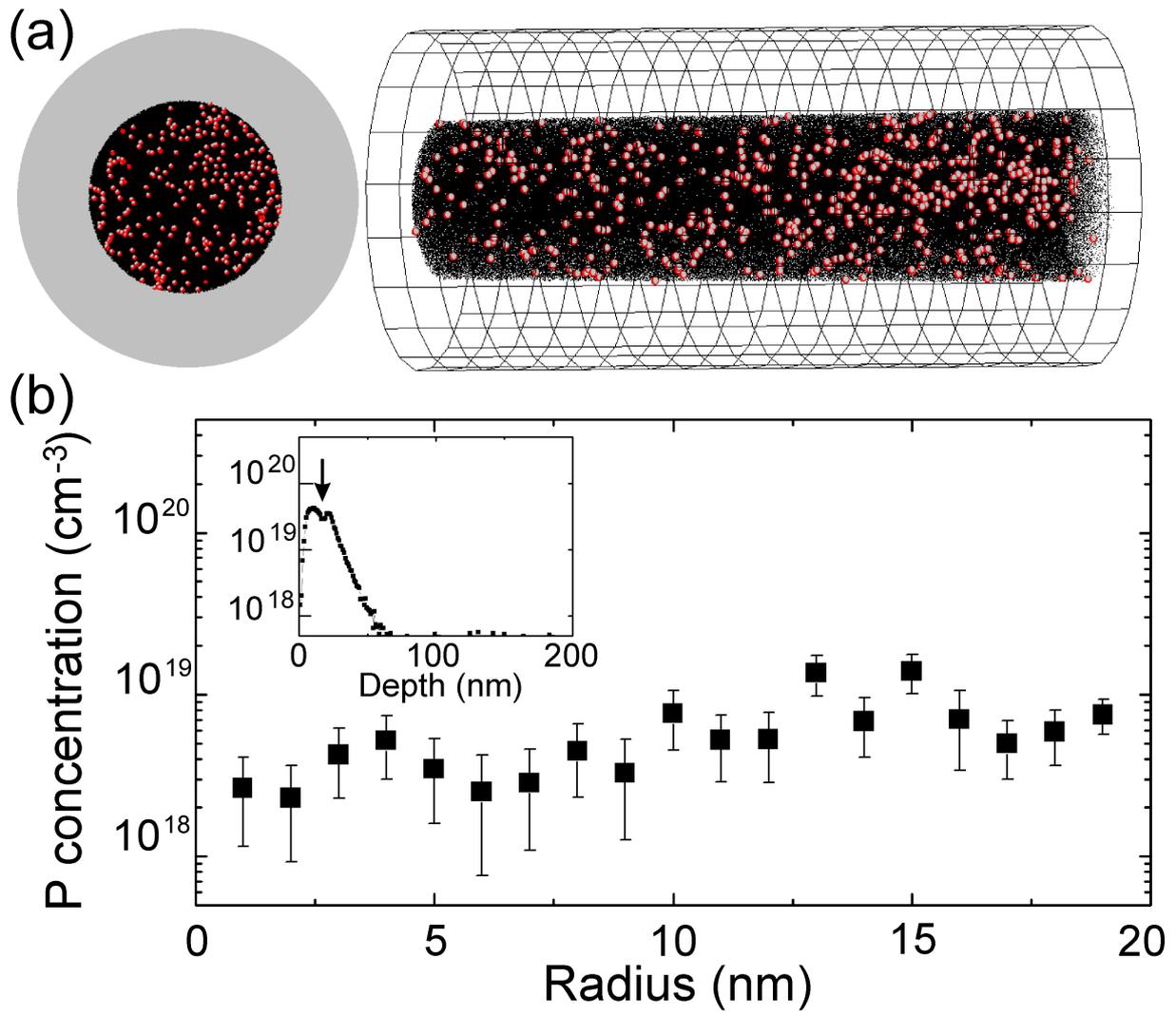

(b)

Figure 3

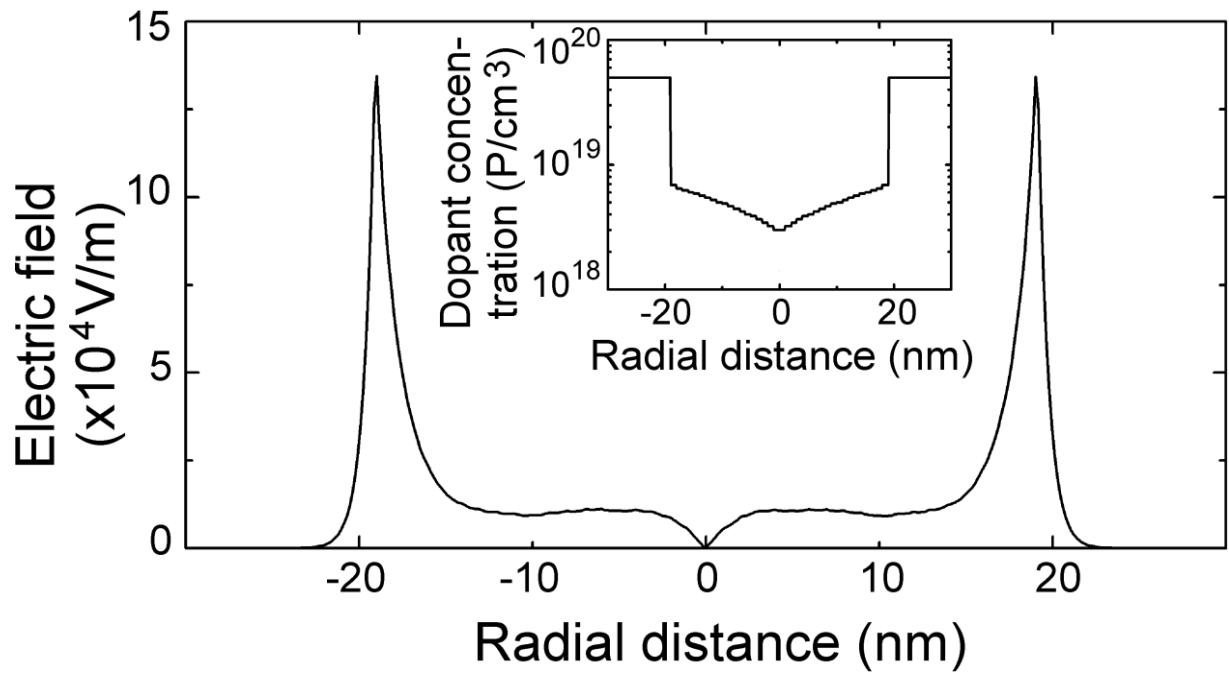

Figure 4

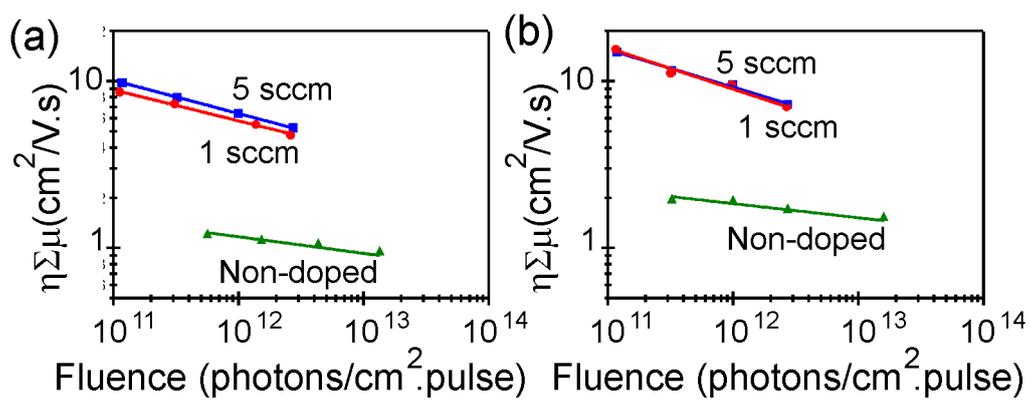

Figure 5